\documentclass[sigconf]{acmart}
\AtBeginDocument{%
  }

\copyrightyear{2026}
\acmYear{2026}
\setcopyright{cc}
\setcctype{by}
\acmConference[KDD '26]{Proceedings of the 32nd ACM SIGKDD Conference on Knowledge Discovery and Data Mining V.1}{August 09--13, 2026}{Jeju Island, Republic of Korea}
\acmBooktitle{Proceedings of the 32nd ACM SIGKDD Conference on Knowledge Discovery and Data Mining V.1 (KDD '26), August 09--13, 2026, Jeju Island, Republic of Korea}
\acmPrice{}
\acmDOI{10.1145/3770854.3783918}
\acmISBN{979-8-4007-2258-5/2026/08}

\usepackage{booktabs}
\usepackage{multirow}
\usepackage{algorithm}
\usepackage{algorithmic}
\usepackage{enumitem}
\usepackage{enumerate}
\usepackage{xcolor}
\usepackage{graphicx}
\usepackage{subcaption}
\usepackage{diagbox}

\newcommand{\eg}{\emph{e.g.}}
\newcommand{\ie}{\emph{i.e.}}

\begin{document}

\title{Sell It Before You Make It: Revolutionizing E-Commerce with Personalized AI-Generated Items}

\author{Jianghao Lin}
\email{linjianghao@sjtu.edu.cn}
\affiliation{
  \institution{Antai College of Economics and Management\\Shanghai Jiao Tong University}
  \city{Shanghai}
  \country{China}
}
\author{Peng Du}
\email{hongyu.dp@taobao.com}
\affiliation{
  \institution{Alibaba Group}
  \city{Hangzhou}
  \country{China}
}
\author{Jiaqi Liu}
\email{1qaz2wsx3edc@sjtu.edu.cn}
\affiliation{
  \institution{Shanghai Jiao Tong University}
  \city{Shanghai}
  \country{China}
}
\author{Weite Li}
\email{ruosong.lwt@taobao.com}
\affiliation{
  \institution{Alibaba Group}
  \city{Hangzhou}
  \country{China}
}
\author{Yong Yu}
\email{yyu@sjtu.edu.cn}
\affiliation{
  \institution{Shanghai Jiao Tong University}
  \city{Shanghai}
  \country{China}
}
\author{Weinan Zhang}
\authornote{Weinan Zhang is the corresponding author.}
\email{wnzhang@sjtu.edu.cn}
\affiliation{
  \institution{Shanghai Jiao Tong University}
  \city{Shanghai}
  \country{China}
}
\author{Yang Cao}
\email{yinming.cy@taobao.com}
\affiliation{
  \institution{Alibaba Group}
  \city{Hangzhou}
  \country{China}
}

\renewcommand{\shortauthors}{Jianghao Lin et al.}

\begin{abstract}
E-commerce has revolutionized retail, yet its traditional workflows remain inefficient, with significant resource costs tied to product design and inventory.
This paper introduces a novel system deployed at Alibaba that uses AI-generated items (AIGI) to address these challenges with personalized text-to-image generation for e-commerce product design. AIGI enables an innovative business mode called ``sell it before you make it'', where merchants can design fashion items and generate photorealistic images with digital models based on textual descriptions. Only when the items have received a certain number of orders, do the merchants start to produce them, which largely reduces reliance on physical prototypes and thus accelerates time to market. 
For such a promising application, we identify the underlying key scientific challenge, \ie, capturing users’ group-level personalized preferences towards multiple generated images. 
To this end, we propose a \underline{Per}sonalized Group-Level Preference Alignment Framework for Dif\underline{fusion} Models (\textbf{PerFusion}). 
We first design PerFusion Reward
Model for user preference estimation with a feature-crossing-based personalized plug-in. 
Then we develop PerFusion with a personalized adaptive network to model diverse preferences across users, and meanwhile derive the group-level preference optimization objective to model comparative behaviors among multiple images.
Both offline and online experiments demonstrate the effectiveness of our
proposed algorithm. 
The AI-generated items achieve
over 13\% relative improvements for both click-through rate and conversion rate, as well as 7.9\% decrease in return rate, compared to their human-designed counterparts, validating the transformative potential of AIGI for e-commerce platforms. 
\end{abstract}




\maketitle

\begin{figure}
  \includegraphics[width=0.45\textwidth]{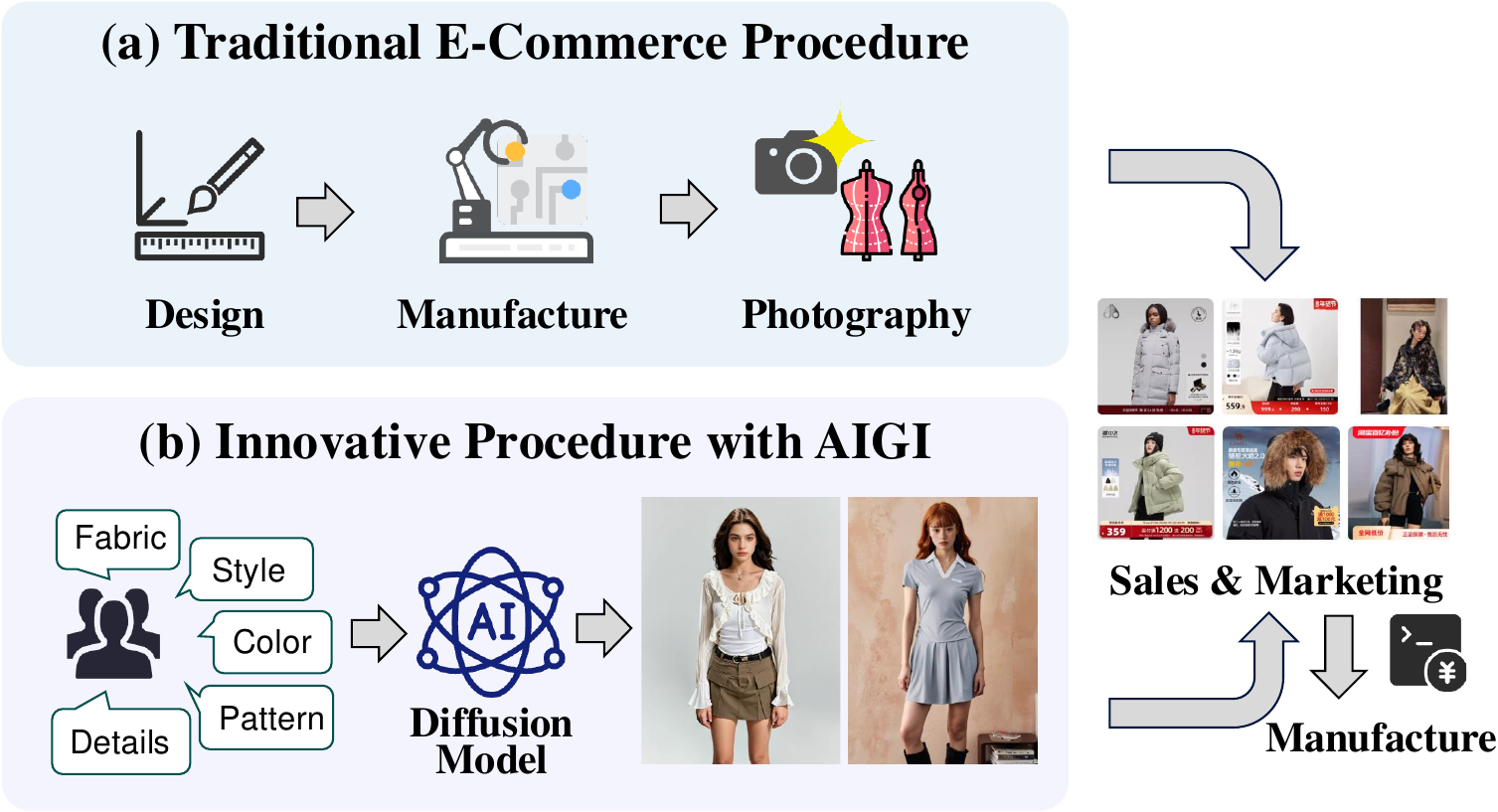}
    \vspace{-6pt}
  \caption{The illustration of (a) traditional procedure and (b) innovative procedure with AI-generated items (AIGI) for e-commerce platforms.}
  \vspace{-5.5mm}
  \label{fig:illustration}
\end{figure}

\section{Introduction}
\label{sec:introduction}

E-commerce has transformed how the world shops, yet its operational processes remain costly and inefficient. 
As shown in Figure~\ref{fig:illustration}(a), traditional workflows demand merchants to design products, manufacture them, and conduct professional photography before listing them online. 
Each step incurs significant time, resource expenditure, and financial risk, particularly when inventory must be committed before gauging consumer demand. 
These inefficiencies not only delay time to market, but also create barriers to innovation and responsiveness in retail.

Recent advancements in generative foundation models provide an opportunity to fundamentally reimagine this process. 
As shown in Figure~\ref{fig:illustration}(b), by leveraging diffusion models, we can automate the product design phase and enable the creation of high-quality and photorealistic visuals based solely on textual descriptions or optional reference images. 
We name it AI-generated items (AIGI).
It not only streamlines the workflow, but also introduces a paradigm shift that represents a fundamental and intrinsic change to the very nature of e-commerce. 
Instead of being constrained by the streamlining of design, manufacturing, and marketing, products can now be designed, marketed, and sold before they are physically manufactured.
This fundamentally alters how supply chains operate, reduces time to market, and minimizes inventory risks, with better flexibility and responsiveness to consumer needs.

\begin{figure}

    \includegraphics[width=0.46\textwidth]{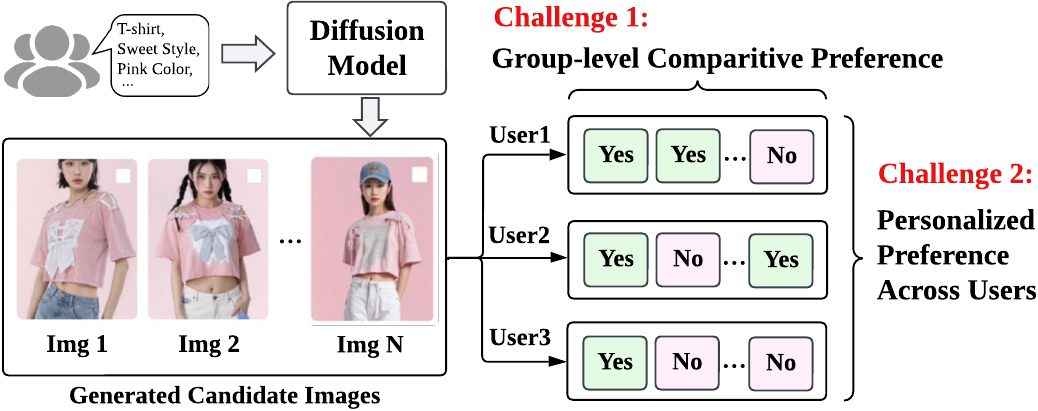}
    \vspace{-6pt}
  \caption{The illustration of general workflow for our online services, as well as the two key challenges: \textit{group-level comparative preference} and \textit{personalized preference across users}.}
  \vspace{-5mm}
  \label{fig:workflow}
\end{figure}

This paper introduces a system deployed at Alibaba that presents a pioneering attempt at realizing the transformative potential of generative AI in the e-commerce industry. 
By integrating cutting-edge generative text-to-image models, we automate the fashion product design process and validate the innovative ``sell it before you make it'' business mode in a real-world e-commerce environment. 
As shown in Figure~\ref{fig:workflow}, the core workflow of our system begins with the user (\ie, the merchant or designer) providing a textual prompt about the desired product design. 
Once the prompt is submitted, a diffusion model generates a diverse set of candidate images. The user then evaluates these candidates, selecting $K$ preferred images from the $N$ generated options. If none of the images meet the user's expectations, the user can refine the text prompt and initiate another round of generation. 
This iterative process empowers users to progressively refine the item designs to align with their specific tastes. 
It is crucial to note that our system is a \textbf{merchant-facing} tool; the end customers on the e-commerce platform are generally \textit{unaware} of whether a product is AI-generated, ensuring a seamless shopping experience.

At the heart of this general workflow lies a critical scientific challenge: \textit{generating realistic product images that best align with users' \textbf{group-level personalized preferences}}. 
As shown in Figure~\ref{fig:workflow}, we decompose it into (1) group-level comparative preference and (2) personalized preference across users.

Firstly, group-level comparative preference means that \textit{user preferences are expressed by comparing a group of images rather than through individual or pairwise evaluations}. 
As shown in Figure~\ref{fig:workflow}, each user should assess the entire group of candidate images holistically to make the decision. 
However, existing preference alignment algorithms for diffusion models either solely rely on positive images without adequately considering negatives~\cite{ho2020denoising}, or focus on pairwise optimization based on one pair of positive and negative images at a time~\cite{wallace2024diffusion,croitoru2024curriculum,liang2024step}. 
These methods are not well suited for our application and would lose the global view of the group-level context, thereby leading to suboptimal performance.

Secondly, personalized preference across users indicates that \textit{user preferences exhibit significant variability, even when the same textual prompt and image set are provided}. This indicates that textual prompts alone are insufficient to fully encapsulate the implicit visual preferences of individual users. 
There are preliminary works exploring user-level preference optimization by incorporating either textual reconstruction~\cite{yang2024new,czapp2024dynamic} or personalized classifier-free guidance with delicately designed networks~\cite{xu2024difashion,xu2024personalized}.
However, they are generally built to re-paint or re-design the reference image based on certain existing contents, \eg, product background outpainting~\cite{shilova2023adbooster,yang2024new}, poster re-painting~\cite{xu2024personalized,shen2024pmg}, and fashion collocation~\cite{xu2024difashion}. 
When faced with the task of generating realistic images from scratch, capturing subtle variations in users' visual preferences still remains a significant challenge.

To this end, in this paper, we propose a \underline{Per}sonalized Group-Level Preference Alignment Framework for Dif\underline{fusion} Models (\ie, \textbf{PerFusion}). 
We first design the PerFusionRM (short for PerFusion Reward Model) for personalized user preference estimation based on the CLIP model~\cite{radford2021learning}.
We propose a feature-crossing-based personalized plug-in to extract the user preference representations and inject them into each Transformer layer of CLIP. 
The design of PerFusionRM serves as the preliminary step to verify the fact that the personalized knowledge can truly affect the user's decision for text-to-image generation. 
Moreover, PerFusionRM is also adopted as the reward model to evaluate the performance of personalized text-to-image generation. 
Then, we propose the PerFusion framework based on the stable diffusion~\cite{rombach2022high} architecture for image generation. 
Specifically, we design a personalized adaptive network to guide the diffusion model with additional user preference conditions, capturing the diverse preferences across different individual users. 
Moreover, a group-level preference optimization objective is derived to model the comparative behavior patterns of users towards multiple candidate images. 

The contributions of this paper are as follows:
\begin{itemize}[leftmargin=10pt]
    \item \textit{We are the first to deploy personalized text-to-image generation for industrial-scale e-commerce item design}. Our system enables merchants to design items and generate photorealistic images before manufacturing, fundamentally transforming the product development cycle and accelerating time to market.
    \item \textit{We identify the key scientific challenge underlying such a promising application}, \ie, capturing the users' group-level comparative behaviors towards multiple candidates, and meanwhile modeling the personalized preferences across different individuals. 
    \item \textit{We propose the PerFusion framework to capture the users' group-level personalized preferences.} 
    We first design PerFusion Reward Model for user preference estimation with a feature-crossing-based personalized plug-in. 
    Then we develop PerFusion based on stable diffusion~\cite{rombach2022high} with a personalized adaptive network to model diverse preferences across users, and meanwhile derive the group-level preference optimization objective to capture the comparative behaviors among multiple candidate images.
    \item \textit{Both offline and online experiments show the effectiveness of our proposed algorithm.}
    This validates the revolutionary potential of the ``sell it before you make it'' business mode based on AI-generated items (AIGI). 
\end{itemize}

\section{Related Works}

\textbf{Image Generation for E-Commerce.}
Image generation in e-commerce aims to customize item displays based on user preferences and maintain visual consistency to enhance the user experience~\cite{lin2025diffusion}. 
Generative models offer automated and personalized content creation of product images, thumbnails, background visuals, and more~\cite{yu2019personalized,lin2024survey,wei2025diffusion,hu2024fashionr2r,kim2024stableviton,vashishtha2024chaining,wang2023generative}. 
These advancements have expanded the application of image generation in e-commerce to areas such as fashion generation and advertisement creative generation.
Fashion generation~\cite{kang2017visually,shih2018compatibility,yang2018recommendation,xu2024diffusion,rombach2022high} involves generating fashion-related contents for personalized fashion recommendations and outfit collocation. 
Advertisement creative generation~\cite{kanungo2021ad,mita2023camera,shilova2023adbooster,yang2024new,czapp2024dynamic,vashishtha2024chaining} focuses on producing engaging advertising content, like promotional fonts, backgrounds, and banners tailored for different audiences. 


\vspace{5pt}
\noindent\textbf{Preference Alignment for Diffusion Models.}
The preference alignment for diffusion models can be categorized into (1) general preference alignment and (2) personalized preference alignment.
\textit{General preference alignment}~\cite{karthik2024scalable,wallace2024diffusion,liang2024step,ho2020denoising} optimizes the text-image consistency and common aesthetic appeal, helping models adhere to visual principles like realism, coherence, and alignment with textual prompts. 
While effective, these approaches only concentrate on the general human preference annotated by crowdsourcing or general reward models, and thereby fail to capture the nuanced preferences of different individual users. 
Hence, personalized preference alignment~\cite{xu2024diffusion,mukande2024mmcrec,czapp2024dynamic,xu2024personalized,wei2025diffusion} emerges as a promising solution, especially for online services like e-commerce.
It adapts generated outputs to individual aesthetic tastes, which are often implicit and difficult to express via text prompts. 


\section{Preliminaries}

\subsection{Problem Formulation}
\label{sec:problem formulation}

We aim to model users' group-level personalized preferences during the image generation process for our e-commerce platforms. 
Specifically, we consider a user, \textbf{typically an e-commerce merchant in our application}, who has a set of $F$ features denoted as $\mathcal{F}=\{f_i\}_{i=1}^F$, \eg, shop style and price level. 
At each round of image generation, the user inputs the text description $txt$ and is provided with a group of generated candidate images $\mathcal{X}=\{x_i\}_{i=1}^N$. 
Then, the user gives the feedback $\mathcal{Y}=\{y_i\}_{i=1}^N$, where $y_i\in\{1,0\}$ indicates whether the user selects (\ie, prefers) the image $x_i$ or not. 
Note that each group of images is represented to the user at the same time, thereby showcasing the group-level comparative preference of the user. 
We assume that different image groups have the same group size of $N$. 
Given this setup, we decompose the goal of modeling users' group-level personalized preferences into two folds:
\begin{itemize}[leftmargin=10pt]
    \item \textbf{User Preference Estimation}: Given a textual instruction $txt$ and a group of candidate images $\mathcal{X}=\{x_i\}_{i=1}^N$, we aim to accurately predict the user's preference for each image within the group, \ie, $\{\hat{y}_{i}\}_{i=1}^N$. 
    We should not only measure the consistency between the textual prompt and generated images, but also capture the user's visual preference among the image group.
    \item \textbf{Personalized Image Generation}: Given a textual instruction $txt$, we aim to generate an image that best aligns with the user’s preferences, \ie, $\hat{x}$. 
    The generated image should not only reflect the given description but also meet the implicit visual preferences inferred from the user's profile features.
\end{itemize}

\subsection{Diffusion Models}
\label{sec:preliminary diffusion}

Our method is built based on stable diffusion~\cite{rombach2022high} (SD).
SD is a latent diffusion model (LDM) that generates high-fidelity images by iteratively refining Gaussian noise through a denoising process. 
Unlike conventional diffusion models operating directly in pixel space~\cite{ho2020denoising}, SD performs diffusion in a latent space, necessitating an autoencoder comprising an encoder and a decoder. 

We denote the encoder as $\mathcal{E}(\cdot)$, which maps an image $x$ to its latent representation $z = \mathcal{E}(x)$. 
Conversely, the decoder, denoted as $\mathcal{D}(\cdot)$, reconstructs the image from its latent representation. Given an input text prompt $txt$, the text encoder $\mathcal{C}(\cdot)$ encodes it into a text embedding $c=\mathcal{C}(txt)$, which serves as the conditioning input for stable diffusion.
During training, for a given diffused latent representation $z_t$ at timestep $t$ with condition $c$, the denoising network $\epsilon_{\theta}(\cdot)$ is trained to predict the noise $\epsilon$. This is achieved by minimizing the following objective function:
\begin{equation}
\mathcal{L}=
\mathbb{E}_{
z \sim \mathcal{E}(x), 
t\sim[0,T], 
c\sim\mathcal{C}(txt), 
\epsilon \sim \mathcal{N}(0,1)}
\left[\left\|\epsilon-\epsilon_\theta\left(z_t, t, c\right)\right\|_2^2\right],
\label{eq:diffusion loss}
\end{equation}
where $\epsilon_{\theta}(\cdot)$ can be typically implemented using the U-Net~\cite{ronneberger2015u} or Transformer~\cite{vaswani2017attention} architecture.

\section{Methodology}


\subsection{Overview of PerFusion}

As discussed in Section~\ref{sec:problem formulation}, the methodology of this paper can be divided into two folds to address the user preference estimation and personalized image generation, respectively. 

We first design the \textbf{PerFusionRM} (short for PerFusion Reward Model) for personalized user preference estimation based on the CLIP model~\cite{radford2021learning,cherti2023reproducible}. 
Specifically, we propose a feature-crossing network to extract the user preference knowledge from the profile features, and then inject the knowledge into each Transformer layer of the CLIP model. 
The feature-crossing network serves as a personalized plug-in, which can retain the general preference estimation of the original CLIP model and leverage the personalized information for adaptation.

The design of PerFusionRM serves as the preliminary step to verify the fact that the personalized information conveyed through the user profiles can ultimately affect the user's final decision towards the group of generated candidate images. 
Moreover, PerfusionRM, which can reflect the user's personalized preference, is also regarded as the reward model to further estimate the performance of diffusion models for personalized image generation tasks.

Then, we introduce the \textbf{PerFusion} framework to diffusion models for personalized image generation. 
To be specific, we adopt the similar feature-crossing network plug-in from the PerFusionRM to extract the user's preference knowledge, which provides personalized conditions for Stable Model through a ControlNet~\cite{zhang2023adding} manner. 
We also derive the group-level preference optimization objective to capture the comparative behavior patterns.

\subsection{User Preference Estimation}
\label{sec:user preference estimation}

\begin{figure}
    \vspace{-10pt}
    \includegraphics[width=0.385\textwidth]{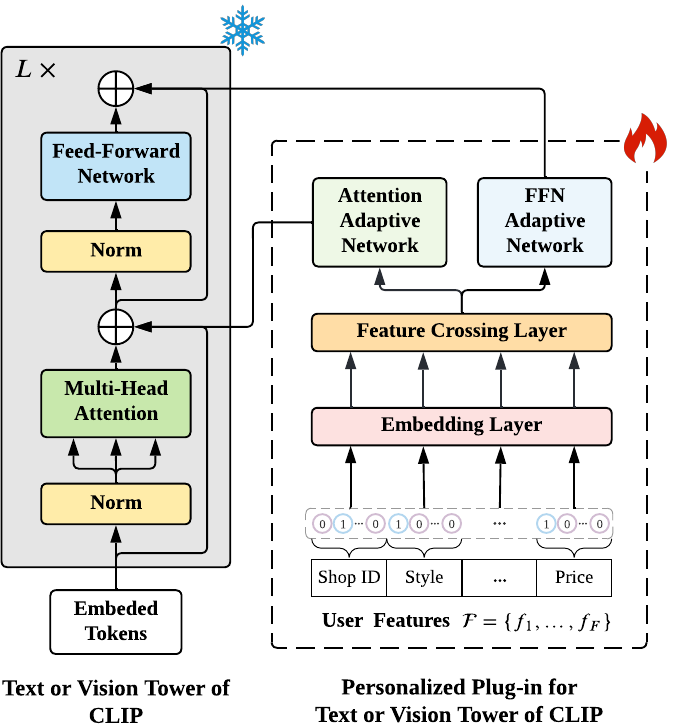}
    \vspace{-7pt}  
  \caption{The architecture of PerFusionRM for user preference estimation based on the pretrained CLIP model. 
  }
    \vspace{-15pt}  
  \label{fig:perfusionRM}
\end{figure}

In this section, we propose PerFusionRM (short for PerFusion Reward Model) for user preference estimation. 
As shown in Figure~\ref{fig:perfusionRM}, it is designed by inserting the personalized plug-in module for the Transformer-based towers of CLIP. 
Since CLIP consists of two Transformer-based towers for text and vision inputs, respectively. 
We would like to first introduce how the personalized plug-in module works for the one tower. 

Specifically, given the user features $\mathcal{F}=\{f_1,\dots,f_F\}$, we first apply the embedding layer to map them to low-dimensional dense embedding vectors $\{v1,\dots,v_F\}$.
Then, following previous research on feature interaction~\cite{wang2021dcn,wang2017deep}, we apply a multi-layer feature crossing network with residual connections to extract the co-occurrence and co-relation patterns among the features:
\begin{equation}
\begin{aligned}
    u_0 &=[v_1 \oplus v_2 \oplus \dots \oplus v_F], \\
    u_{l+1} &= u_0 u_l^T w_l + b_l + u_l,
\end{aligned}
\end{equation}
where $u_l$ is the output of the $l$-th feature-crossing layer, and $\{w_l,b_l\}$ are the learnable parameters.
In this way, we extract the user preference representation from the input features, denoted as $u$.

Each Transformer layer of CLIP can be formulated as a multi-head attention (MHA) followed by a feed-forward network (FFN). We ignore the normalization and residual connections for brevity:
\begin{equation}
\begin{aligned}
    [h_{l,i}^{\prime}]_{i=1}^T=\text{MHA}\left([h_{l,i}]_{i=1}^T\right),\;\;[h_{l+1,i}]_{i=1}^T=\text{FFN}\left([h_{l,i}^{\prime}]_{i=1}^T\right),
\end{aligned}
\end{equation}
where $h_{l,i}$ is the hidden state for the $i$-th token at the $l$-th Transformer layer. 
To integrate the user preference knowledge, we feed the user representation $u$ through the attention adaptive network and FFN adaptive network to obtain the adaptive knowledge representations $u_{\text{attn}}^l$ and $u_{\text{ffn}}^l$, respectively:
\begin{equation}
\begin{aligned}
    u_{\text{attn}}^l=\text{AdaNet}_l(u),\;u_{\text{ffn}}^l=\text{AdaNet}_l(u),\;l=1,\dots,L
\end{aligned}
\end{equation}
where $L$ is the number of Transformer layers for the CLIP tower, and the adaptive networks are implemented as two-layer perceptrons with GeLU activations~\cite{hendrycks2016gaussian}. 
Then, we insert the adaptive knowledge $\{u_{\text{attn}}^l,u_{\text{ffn}}^l\}$ into the Transformer layer with element-wise addition:
\begin{equation}
\begin{aligned}
    [h_{l,i}^{\prime}]_{i=1}^T&=\text{MHA}\left([h_{l,i} \oplus u_{\text{attn}}^l]_{i=1}^T\right),\\
    [h_{l+1,i}]_{i=1}^T&=\text{FFN}\left([h_{l,i}^{\prime} \oplus u_{\text{ffn}}^l]_{i=1}^T\right).
\end{aligned}
\end{equation}
Note that the last linear layer of the adaptive network is initialized with zero parameters to ensure that the initial outputs of adaptive networks at the beginning of training would not affect the CLIP output, \ie, $u_{\text{attn}}^l=u_{\text{ffn}}^l=\mathbf{0}$.

The process introduced above mainly describes how to inject the user preference knowledge inferred by the personalized plug-in into one Transformer-based tower of CLIP. 
We can simply duplicate or share the personalized plug-in for the two towers of CLIP. 
In practice, we suggest duplication instead of sharing, since the knowledge required for textual and visual knowledge extraction can conflict with each other and lead to suboptimal performance. 

In this way, PerFusionRM is able to capture the nuanced individual preference and meanwhile preserve the general preference estimation for text-image pairs. 
We can infer the representations $(v_{\text{text}}^u,v_{\text{image}}^u)$ for a given text-image pair $(txt, x)$ based on the two-tower CLIP model with personalized plug-ins, and then perform the user preference estimation via the cosine similarity:
\begin{equation}
    s(txt,x)=\text{Cosine\_Similarity}(v_{\text{text}}^u,v_{\text{image}}^u).
\end{equation}

Each training sample can be formulated as $\{txt, \mathcal{X}, \mathcal{F}, \mathcal{Y}\}$, where $txt$ is text description, $\mathcal{X}=\{x_i\}_{i=1}^N$ is the group of candidate images, $\mathcal{F}$ is the user feature set, and $\mathcal{Y}=\{y_i\}_{i=1}^N$ is the set of corresponding binary labels. 
We can compute the ideal preference distribution as:
\begin{equation}
    p_{i}=y_i \;/\; \sum\nolimits_{j=1}^N y_j,\;\,i=1,\dots,N.
\end{equation}
For instance, if the user prefers the first two images and rejects all the rest, then the ideal preference distribution is $[0.5, 0.5, 0,\dots]$.
Similarly, the predicted distribution can be calculated as:
\begin{equation}
    \hat{p}_{i}=\frac{\exp(s(txt,x_i))}{\sum_{j=1}^N\exp(s(txt,x_j)))},\;\,i=1,\dots,N.
\end{equation}
Hence, the group-wise training objective can be written as:
\begin{equation}
    \mathcal{L}_{RM}=-\sum\nolimits_{i=1}^N p_{i}\log \hat{p}_{i}+(1-p_{i})\log(1-\hat{p}_{i}).
    \label{eq:total loss}
\end{equation}
This objective enables the injection of rich, structured merchant-side information into the reward model, enabling fine-grained preference learning.
Note that we freeze the original CLIP backbone and solely finetune the parameters of personalized plug-in modules, which helps retain the basic knowledge for estimating text-image pairs.

\subsection{Personalized Image Generation}

We propose the PerFusion framework for personalized group-level preference alignment based on stable diffusion~\cite{rombach2022high}. 
As shown in Figure~\ref{fig:perfusion}, we first design the personalized adaptive network to capture the diverse preference across different users, and then derive the group-level preference optimization objective to model the users' comparative pattern among multiple candidate images. 

\subsubsection{Personalized Adaptive Network}

\begin{figure}
  \includegraphics[width=0.39\textwidth]{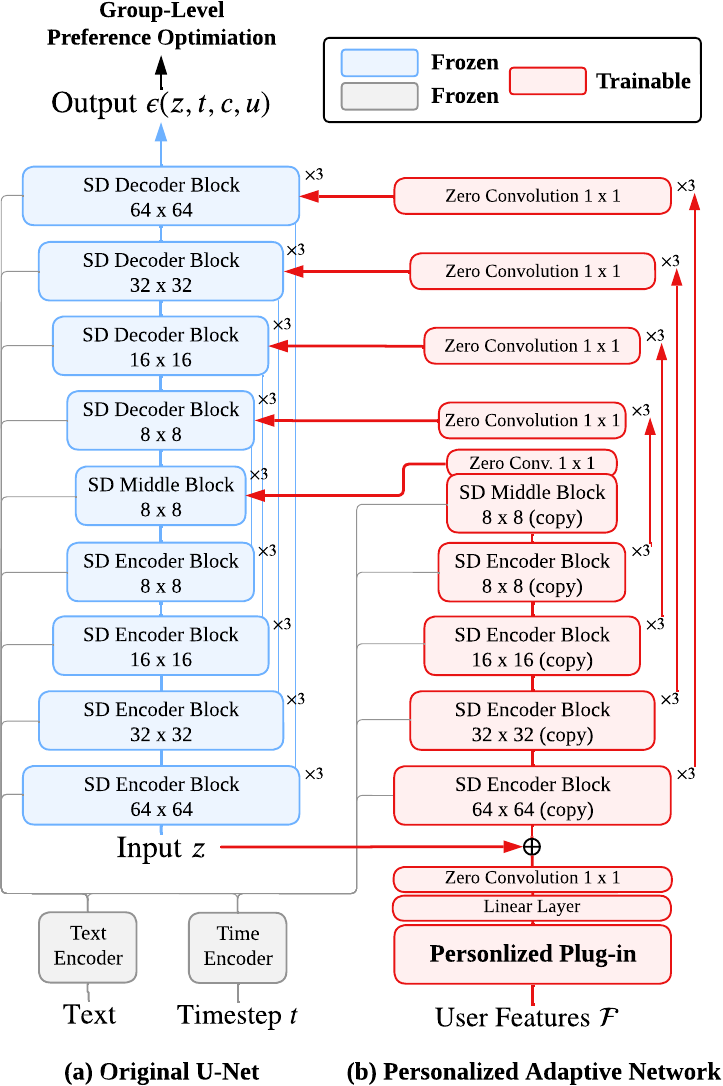}
    \vspace{-7pt}
  \caption{The overall framework of PerFusion built based on the U-Net architecture. 
  The left part is the classical U-Net model, and the right part is the adaptive network associated with the personalized plug-in introduced in Section~\ref{sec:user preference estimation}.}
    \vspace{-10pt}
  \label{fig:perfusion}
\end{figure}

Stable diffusion is essentially a U-Net~\cite{ronneberger2015u} with an encoder, a middle block, and a skip-connected decoder. 
As shown in Figure~\ref{fig:perfusion}(a), it contains a total of 25 blocks, including one middle block, and 12 blocks for encoder and decoder, respectively. 
Each block consists of convolution layers and other necessary neural operators like pooling and batch normalization.
The symbol ``$\times3$'' indicates that the corresponding block repeats three times. 
The U-Net takes the latent representation $z_t$ to be denoised, the timestep $t$, and the textual condition $c$ as inputs, and outputs the noise to be eliminated, \ie, $\epsilon_{\theta}(z_t,t,c)$. 

As for personalized adaptive network shown in Figure~\ref{fig:perfusion}(b), we first adopt the feature-crossing-based personalized plug-in introduced in Section~\ref{sec:user preference estimation} to extract the user representation $u$. 
We conduct the following operations to obtain the input for adaptive network:
\begin{itemize}[leftmargin=10pt]
    \item Apply a linear layer to $u$ and reshape the output into a 2D feature map with the same size as the input latent $z_t$;
    \item Perform $1\times1$ convolution layer with both weight and bias initialized to zeros;
    \item Do element-wise addition with $z_t$ to integrate the denoising information with user preference, resulting in $z_t^{\prime}$.
\end{itemize}
Then, we create trainable copies of the 12 encoding blocks and 1
middle block of U-Net. 
The integrated input $z_t^{\prime}$, together with the textual condition $c$ and timestep $t$, is passed through these trainable copy blocks to extract the essential preference knowledge needed in this round of denoising. 
We feed the output of each trainable copy block to the corresponding $1\times1$ zero convolution layer, and finally add it back to the decoder or middle blocks in the original U-Net. 
Note that the zero convolution means that the weight and bias are all initialized to zeros, which protects the backbone U-Net from the random noise in the initial training steps. 

Putting it all together, we are able to predict the noise to be eliminated at timestep $t$ by further incorporating the user's personalized preference knowledge, \ie, $\epsilon_{\theta}(z_t,t,c,u)$.

\subsubsection{Group-Level Preference Optimization}

We derive the training objective based on DPO~\cite{wallace2024diffusion,rafailov2024direct}, the most popular pairwise preference alignment algorithm.
Formally, we can divide the candidate image set $\mathcal{X}$ into a positive set $\mathcal{P}$ and a negative set $\mathcal{N}$ according to user binary feedback. 
Then, we can first substitute the pairwise Bradley-Terry model~\cite{bradley1952rank} in DPO~\cite{rafailov2024direct} with the Plackett-Luce model~\cite{luce1959individual,plackett1975analysis} to consider group-level preference distributions with an independent assumption among positive samples:
\begin{equation}
\begin{aligned}
    &p^{*}(x_p  \succ x_n,\forall x_p\in\mathcal{P},\forall x_n\in\mathcal{N}) \\
    =\;&\sum\nolimits_{x_p\in\mathcal{P}}p(x_p  \succ x_n,\forall x_n\in\mathcal{N}) \;\;\; \text{(by independent assumption)} \\
    =\;&\sum\nolimits_{x_p\in\mathcal{P}}\frac{1}{1+\sum_{x_n\in\mathcal{N}}\exp(r(x_n,c)-r(x_p,c))},
\end{aligned}
\label{eq:preference distribution}
\end{equation}
where $r(x,c)$ is the reward model that can be derived according to previous works~\cite{wallace2024diffusion,rafailov2024direct} with a fixed reference model $p_{ref}(\cdot)$:
\begin{equation}
    r(x,c)=\beta \log\frac{p_{\theta}(x|c)}{p_{ref}(x|c)}+\beta\log Z(c).
\end{equation}
Hence, by substituting the reward function into Eq.~\ref{eq:preference distribution}, we can then derive the group-level training objective:
\begin{equation}
\begin{aligned}
    \mathcal{L}=&-\mathbb{E}_{(u,c,\mathcal{P},\mathcal{N})}\left[\log p^{*}(x_p  \succ x_n,\forall x_p\in\mathcal{P},\forall x_n\in\mathcal{N})\right]\\
    = &-\mathbb{E}_{(u,c,\mathcal{P},\mathcal{N})}\Bigg[\sum_{x_p\in\mathcal{P}} \log\sigma\bigg(-\log\sum_{x_n\in\mathcal{N}}\exp\Big(\\
    &\;\;\;\;\;\;\;\;\;\beta \log\frac{p_{\theta}(x_n|c)}{p_{ref}(x_n|c)}-\beta \log\frac{p_{\theta}(x_p|c)}{p_{ref}(x_p|c)}\Big) \bigg)
    \Bigg].
\end{aligned}
\end{equation}
However, it is intractable to calculate the likelihood of an image, \ie, $p_{\theta}(x|c)$. Therefore, following previous work~\cite{wallace2024diffusion}, we approximate the denoising trajectory $p_{\theta}(x|c)$ with the diffusion process $q(x|c)$ and derive the final objective:
\begin{equation}
\small
    \mathcal{L}= -\mathbb{E}_{(u,t,c,\mathcal{P},\mathcal{N})} \sum_{x_p\in\mathcal{P}} \log\sigma\bigg(-\log\sum_{x_n\in\mathcal{N}}\exp\Big(\beta s(x_p,c,t,u)-\beta s(x_n,c,t,u)
    \Big) \bigg),
    \label{eq:group-level objective}
\end{equation}
where $s(x,c,t,u)=\|\epsilon^* - \epsilon_\theta(z_t,t,c,u)\|^2_2 - \|\epsilon^* - \epsilon_\text{ref}(z_t,t,c,u)\|^2_2$, $u$ is the user preference representation, $z_t$ is the latent diffused representation of image $x$, and $\beta$ is a hyper-parameter for regularization. 
We give detailed derivation and analysis in Appendix~\ref{app:loss derive}.
\section{Offline Experiment}

\subsection{Experiment Setup}
\subsubsection{Datasets}
We adopt the public PickaPic dataset~\cite{kirstain2023pick} and Alibaba's Industrial dataset for experiments. 
We show the dataset statistics in Table~\ref{tab:datasets}.

\begin{table}[h]
    \vspace{-3pt}
    \caption{The dataset statistics.}
    \vspace{-7pt}
    \centering
    \resizebox{0.4\textwidth}{!}{
    \renewcommand\arraystretch{1.08}
    \begin{tabular}{c|ccccc}
    \toprule
     Dataset   & \#Users & \#Prompts & \#Samples & Group Size \\ 
     \midrule
     PickaPic  & 3,094 & 33,164 & 529,219 & 2 \\
     Industrial & 54,995 & 60,616 & 607,328 & 5 \\
     \bottomrule
    \end{tabular}
    }
    \vspace{-7pt}
    \label{tab:datasets}
\end{table}

Crucially, while Pickapic dataset lacks the rich merchant-specific features of our industrial data, it provides a user ID for each preference judgment. This allows our personalized models to learn user-specific tastes, although the more limited feature set explains the comparatively smaller performance gains observed on this dataset.

We filter Pickapic and industrial datasets to ensure each user has at least 10 group-level records.
We split datasets into train/valid/test sets with 8:1:1 ratio at group record level, and all evaluations are done on hold-out sets.


\subsubsection{Baselines}
For \textit{user preference estimation} task, we adopt Aesthetics~\cite{aesthetic_classifier}, CLIP-H~\cite{radford2021learning}, ImageReward~\cite{xu2024imagereward}, HPSv1~\cite{wu2023human}, HPSv2~\cite{wu2023humanv2} and PickScore~\cite{kirstain2023pick} as the baselines. 
For \textit{personalized image generation} task, we adopt Stable Diffusion v1.5~\cite{rombach2022high}, SFT~\cite{ho2020denoising}, DPO~\cite{wallace2024diffusion}, SPO~\cite{liang2024step}, PMG~\cite{shen2024pmg} as baselines.
We give detailed information about baselines in Appendix~\ref{app:baseline models} due to page limitations.

\subsubsection{Evaluation Metrics} 
For \textit{user preference estimation} task, we formulate it as a ranking task to rank positive images with higher scores. Hence, we adopt the mean averaged precision (MAP) and group-wise area under the curve (GAUC) as the metrics.
For the \textit{personalized image generation} task, following previous works~\cite{wallace2024diffusion, ho2020denoising, liang2024step, xu2024personalized} we use Aesthetic~\cite{aesthetic_classifier}, CLIP Score~\cite{radford2021learning}, and PickScore~\cite{kirstain2023pick} to evaluate the general quality of generated images. 
Moreover, we adopt our proposed PerFusionRM as the reward model to estimate whether the generated images can satisfy the users' personalized preferences. 
We give detailed information for evaluation metrics in Appendix~\ref{app:evaluation metrics} due to page limitations.

\subsubsection{Implementation Details}
We give the implementation details in Appendix~\ref{app:implementation} due to page limitations.

\begin{table}
\caption{The comparison of different reward models for user preference estimation task.  The best results are given in bold, while the second-best results are underlined.}

\vspace{-10pt}
\label{tab:Reward Model}
\resizebox{0.39\textwidth}{!}{
\renewcommand\arraystretch{1.2}
\begin{tabular}{c|cc|cc}
\toprule
\hline

\multicolumn{1}{c|}{\multirow{2}{*}{Reward Model}} & \multicolumn{2}{c|}{PickaPic} & \multicolumn{2}{c}{Industrial}  \\ 
\multicolumn{1}{c|}{} & MAP $\uparrow$  & GAUC $\uparrow$ & MAP $\uparrow$ & GAUC $\uparrow$ \\ 
   \hline 
Aesthetics & 0.7569 & 0.5138 & 0.4571 & 0.5000  \\ 
CLIP-H & 0.8148 & 0.6295 & \underline{0.6352} & \underline{0.7060}  \\ 
ImageReward & 0.8216 & 0.6431 & 0.5538 & 0.6115  \\ 
HPSv1 & 0.8293 & 0.6585 & 0.4565 & 0.5030 \\ 
HPSv2 & 0.8177 & 0.6353 & 0.5783 & 0.6420  \\ 
PickScore & \underline{0.8711} & \underline{0.7422} & 0.5241 & 0.5820  \\ 
PerFusionRM (Ours) & \textbf{0.8822} & \textbf{0.7644} & \textbf{0.9220} & \textbf{0.9564} \\
Rel.Imprv. & 1.27\% & 2.99\% & 45.15\% & 35.47\% \\
  
   \hline  
   \bottomrule          
\end{tabular}
}
\vspace{-2.5mm}
\end{table}

\subsection{User Preference Estimation}
\label{sec:exp user preference estimation}

We report the performance of user preference estimation task in Table~\ref{tab:Reward Model}, from which we can obtain the following observations:
\begin{itemize}[leftmargin=10pt]
    \item PerFusionRM significantly outperforms all the baseline models on both datasets. 
    The baseline models either focus solely on text-image consistency and aesthetic quality (\eg, CLIP-H) or consider only general human preferences (\eg, PickScore), neglecting personalized preferences across different individual users.
    In contrast, PerFusionRM employs a feature-crossing network to extract the user preference knowledge and inject it into each Transformer layer of the CLIP model, resulting in superior performance for personalized user preference estimation tasks.
    \item The relative improvement of PerFusionRM over the best baseline is significantly larger on Industrial dataset than that on PickaPic dataset. 
    We attribute this phenomenon to the potential difference in data sources between these two datasets:
    \begin{itemize}[leftmargin=10pt]
        \item PickaPic is a public dataset, where the candidate images are generated by various open-sourced models like stable diffusion variants~\cite{rombach2022high}. 
        The images for text prompts might be imperfect and force users to concentrate more on the text-image consistency and general aesthetic quality. 
        Hence, personalized preference factors are weakened on PickaPic dataset.
        \item The images in Industrial dataset are generated by our online serving models, which are well-tuned to satisfy the users' need for product design. 
        The general preference and text-image consistency can be already satisfied for all the generated candidates. 
        As a result, the personalized preference factors are much more important for Industrial dataset, leading to a more significant improvement of PerFusionRM over baselines.
    \end{itemize}
    
    
\end{itemize}

\subsection{Personalized Image Generation}

We report the performance of image generation in Table~\ref{tab:Generate image performance}, from which we can draw the following observations:
\begin{itemize}[leftmargin=10pt]
    \item PerFusion achieves the best performance over all the baseline models, except for a few cases. 
    This validates the effectiveness of PerFusion in not only satisfying the general human preference (\eg, text-image consistency and aesthetic appeal), but also capturing the diverse preferences across different individual users. 
    \item PerFusion can achieve better performance improvement on Industrial dataset than that on PickaPic dataset in terms of the PerFusionRM metric. 
    This indicates that the factor of personalized preferences is more important on Industrial dataset, which is consistent to the results reported in Table~\ref{tab:Reward Model} at Section~\ref{sec:exp user preference estimation}. 
\end{itemize}

\begin{table}
\caption{
Performance comparison for the image generation task.
The best results are given in bold, and the second-best values are underlined. 
}

\vspace{-10pt}
\label{tab:Generate image performance}
\resizebox{0.45\textwidth}{!}{
\renewcommand\arraystretch{1.1}
\begin{tabular}{c|cccc}
\toprule
\hline

\multicolumn{1}{c|}{\multirow{2}{*}{Model}} & \multicolumn{4}{c}{PickaPic} \\ 
\multicolumn{1}{c|}{} & Aesthetic $\uparrow$ & CLIP Score $\uparrow$ & PickScore $\uparrow$ & PerFusionRM $\uparrow$ \\ 
   \hline 
   
SD1.5 & 5.39 & 32.82 & 20.29 & 21.23 \\
SFT & \underline{5.71} & 34.58 & 20.89 & 21.80 \\
SPO & 5.60 & 31.51 & 20.72 & 21.57 \\
DPO & 5.67 & 34.67 & 21.13 & \underline{22.62} \\
PMG & 5.70 & \underline{35.21} & \underline{21.32} & 22.51 \\
PerFusion (Ours) & \textbf{5.80} & \textbf{35.93} & \textbf{21.65} & \textbf{23.03} \\
  
   \hline 

\multicolumn{1}{c|}{\multirow{2}{*}{Model}} & \multicolumn{4}{c}{Industrial} \\ 
\multicolumn{1}{c|}{} & Aesthetic $\uparrow$ & CLIP Score $\uparrow$ & PickScore $\uparrow$ & PerFusionRM $\uparrow$ \\ 
   \hline

SD1.5 & 5.36 & 17.73 & 17.01 & 13.44 \\
SFT & \textbf{5.44} & 24.39 & 18.07 & 24.33 \\
SPO & 5.25 & 19.27 & 14.63 & 17.72 \\
DPO & 5.19 & \underline{25.68} & 18.19 & 25.03 \\
PMG & 5.25 & 25.49 & \underline{18.27} & \underline{25.33} \\
PerFusion (Ours) & \underline{5.41} & \textbf{25.74} & \textbf{18.54} & \textbf{26.44} \\

   \hline
   \bottomrule          
\end{tabular}
}
\end{table}

\subsection{Ablation Study}

In this section, we conduct the ablation study for PerFusionRM and PerFusion on both user preference estimation and personalized image generation tasks, respectively. 

\begin{figure}[t]
    \centering
    \includegraphics[width=\linewidth]{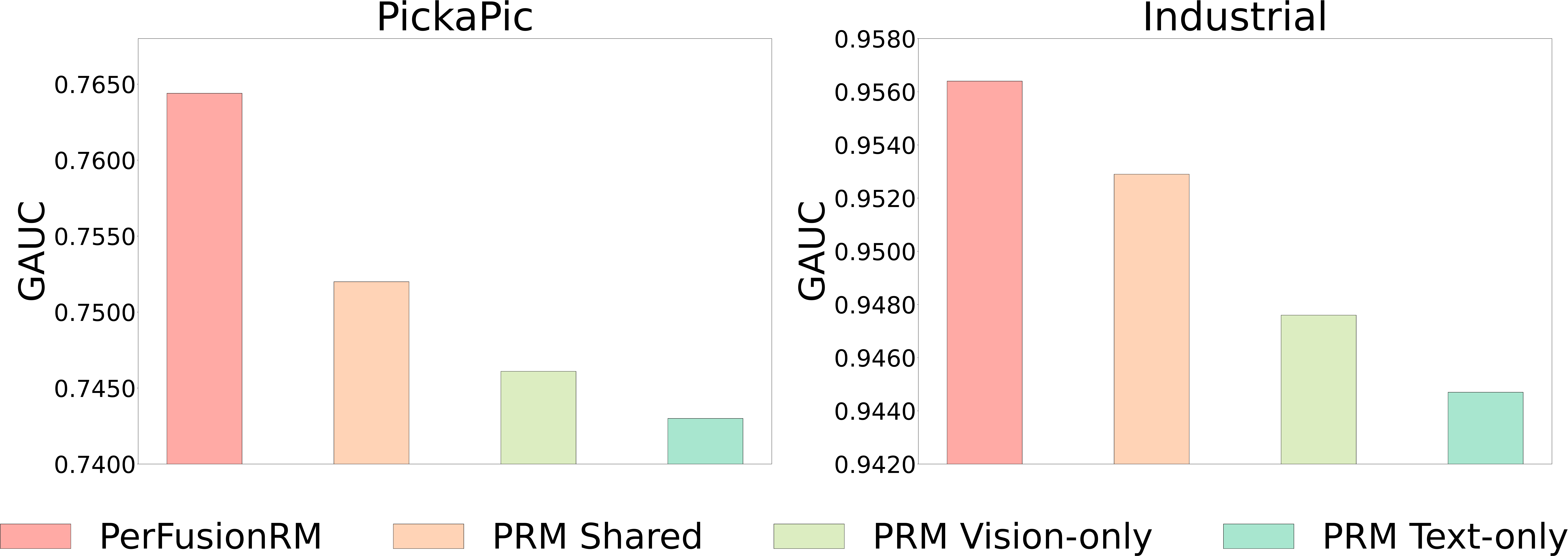}
    \vspace{-20pt}
    \caption{Ablation study of PerFusionRM for the user preference estimation task.}
    \vspace{-10pt}
    \label{fig:reward model ablation}
\end{figure}

For the user preference estimation task, we develop the following model variants:
\begin{itemize}[leftmargin=10pt]
    \item \textbf{PerFusionRM}. The complete version of our proposed model for preference estimation, where we duplicate two independent personalized plug-ins for the text and vision towers in CLIP. 
    \item \textbf{PRM Shared}. We share the same personalized plug-in for the text and vision towers in CLIP instead of duplicating.
    \item \textbf{PRM Vision-Only}. We remove the personalized plug-in for the text tower and keep it for the vision tower. 
    \item \textbf{PRM Text-Only}. We remove the personalized plug-in for the vision tower and keep it for the text tower. 
\end{itemize}
The results are shown in Figure~\ref{fig:reward model ablation}. 
We can observe that either removing or sharing the personalized plug-ins for the text and vision towers in CLIP can lead to performance degradation, indicating that the
knowledge required for textual and visual knowledge extraction is indispensable and distinct from each other.

\begin{figure}[t]
    \centering
    \includegraphics[width=\linewidth]{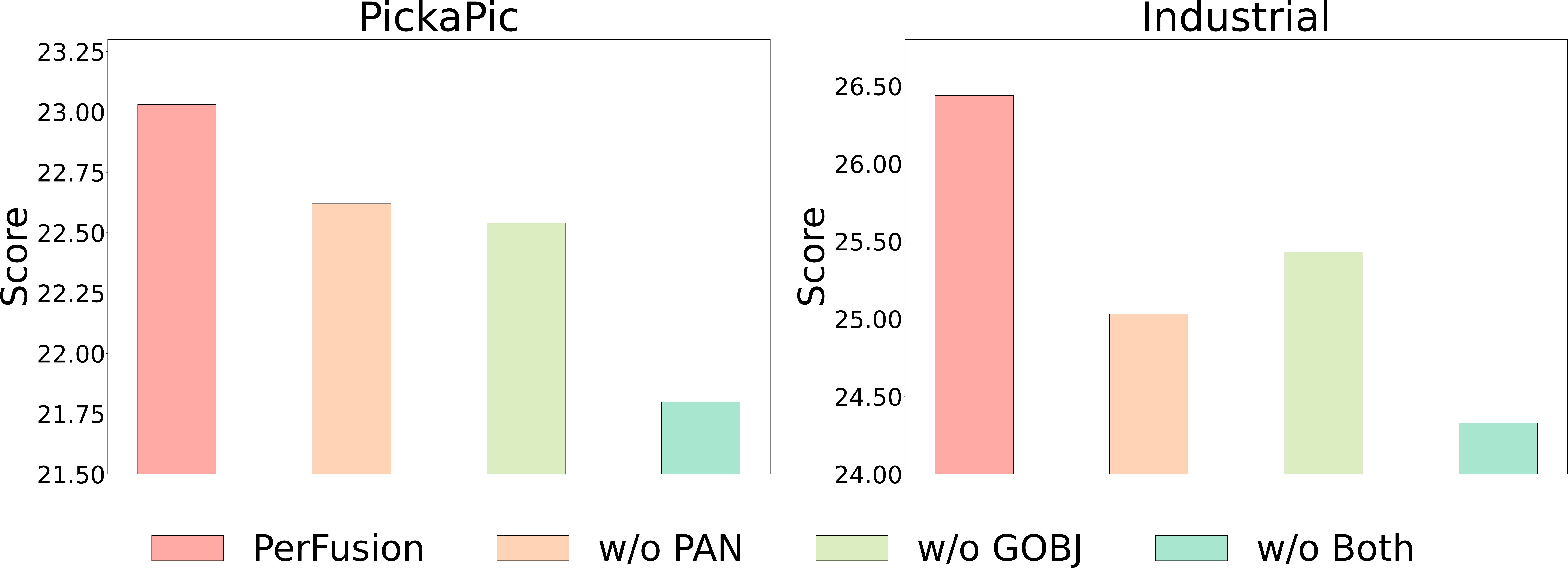}
    \vspace{-20pt}
    \caption{Ablation study of PerFusion for the personalized image generation task.}
    \label{fig:generation ablation}
\end{figure}

For the personalized image generation task, we evaluate the following model variants with the PerFusionRM metric:
\begin{itemize}[leftmargin=10pt]
    \item \textbf{PerFusion}. The full version of our proposed model for personalized image generation.
    \item \textbf{PF w/o PAN}. We remove the personalized adaptive network. 
    \item \textbf{PF w/o GOBJ}. We remove the group-level preference optimization objective and simply train the model by Eq.~\ref{eq:diffusion loss}.
    \item \textbf{PF w/o Both}. We remove both the personalized adaptive network and group-level preference optimization objective.
\end{itemize}
The results are shown in Figure~\ref{fig:generation ablation}. 
                                                                                                                                                                                                                                                                                                                                                                                                            We observe that removing either PAN or GOBJ can degrade the performance for personalized image generation. 
This validates the effectiveness of both personalized adaptive network and group-level preference optimization objective in modeling users' group-level personalized preferences.


\subsection{Computational and Memory Cost}

We train our models on 8 NVIDIA GPUs and perform inference on a single GPU. As shown in Table~\ref{tab:computational cost}, both the training and inference time of PerFusion are higher than the standard diffusion model. This increased overhead is brought by the personalized adaptive network required for processing user-specific features. 
In our production environment, this trade-off is justified by the significant business impact, including a +17.27\% increase in conversion rate and -7.9\% decrease in return rate for the apple-to-apple comparison in the same store.
This makes the additional cost a worthwhile investment for achieving superior personalization.

As for memory consumption, PerFusion runs with FP16 inference, using ~9GB GPU memory (vs. 5GB for SD1.5) due to the personalized adaptive network. In production, we apply techniques like xFormers to further reduce memory usage, making the system scalable for large-scale e-commerce deployment.
\begin{table}[t]
\caption{The computational cost analysis of different models in terms of training and inference time.}
\vspace{-7pt}
\centering
\resizebox{0.41\textwidth}{!}{
\renewcommand\arraystretch{1.08}
\begin{tabular}{c|cc|cc}
\toprule
\hline

\multicolumn{1}{c|}{\multirow{2}{*}{Model}} & \multicolumn{2}{c|}{Train Time/Batch} & \multicolumn{2}{c}{Infer. Time/Sample} \\

\multicolumn{1}{c|}{} & Pickapic      & Industrial      & Pickapic       & Industrial       \\ 

   \hline 

Diffusion            & 1.45s         & 1.48s           & 1.49s          & 1.49s            \\
DiffusionDPO         & 3.05s         & 3.16s           & 1.48s          & 1.50s            \\
PerFusion            & 6.29s         & 6.95s           & 3.03s          & 3.05s            \\
\hline
\bottomrule
\end{tabular}
}
\label{tab:computational cost}
\end{table}

\subsection{Backbone Analysis of PerFusionRM}

We further conduct experiments by replacing the CLIP backbone of PerFusionRM with other multimodal models:
\begin{itemize}[leftmargin=10pt]
    \item \textbf{SigLIP}~\cite{zhai2023sigmoid}: A dual-tower model trained with pairwise sigmoid loss over visual and textual encoders.
    \item \textbf{ViLT}~\cite{kim2021vilt}: A single-tower model that jointly captures the relationship of vision and text based on shallow input layers.
\end{itemize}

\begin{table}[t]
\vspace{-7pt}
\caption{The backbone analysis of PerFusionRM.}
\label{tab:backbone perfusionrm}
\vspace{-7pt}
\centering
\resizebox{0.25\textwidth}{!}{
\begin{tabular}{lccc} 
\toprule
Backbone & CLIP & SigLIP & ViLT \\
\midrule
MAP  & \textbf{0.8822} & 0.8798 & 0.8776 \\
GAUC & \textbf{0.7644} & 0.7601 & 0.7577 \\
\bottomrule
\end{tabular}
}
\vspace{-7pt}
\end{table}
The results are reported in Table~\ref{tab:backbone perfusionrm}. CLIP outperforms SigLIP, even though they both use dual-tower architectures. This is mainly because we adopt the PickScore-style CLIP variant, which is first pre-trained on the same dataset to better fit the distribution, leading to more accurate user preference modeling. ViLT, as a single-tower model that fuses vision and text inputs via concatenation, performs relatively worse. A likely reason is that only one personalized plug-in module can be applied to the mutual text-image encoder, which limits its ability to separately model user preferences over vision and text. In contrast, dual-tower models can disentangle these modalities, allowing more fine-grained and flexible personalization.

\section{Industrial Deployment}

We provide a detailed description and discussion about the real-world deployment of personalized image generation for AI-generated items (AIGI) in Alibaba, a world-wide e-commerce platform. 
Specifically, in our application, the generation \textbf{mainly focuses on the fashion topics}, which can be easily supported for manufacturing solely based on the product design figures.

\subsection{Human Expert Evaluation}

\begin{table}
\vspace{-5pt}
\caption{The human expert evaluation on the image qualities of AI-generated items and human-designed real items from six aspects under a three-tier scoring system, \ie, 0 for poor, 1 for average, and 2 for good.}
\vspace{-10pt}
\label{tab:human expert}
\resizebox{0.48\textwidth}{!}{
\renewcommand\arraystretch{1.2}
\begin{tabular}{c|c|c|c}
\toprule
\hline 
Evaluation Aspects  & AI-generated & Human-designed & AI vs Human Ratio \\ 
\hline 
Material Quality  & 1.5168 & 1.6260 & 93.28\% \\
Model Realism  & 1.7020 & 1.8888  & 90.11\% \\
Craft Design & 1.4520 & 1.7648 & 82.28\% \\
Pattern Appropriateness  & 1.6972 & 1.8464 & 91.92\% \\
Design Aesthetics & 1.7932 & 1.8708 & 95.85\% \\
Color Coordination & 1.7532 & 1.7932 & 97.77\% \\
\hline  
Overall & 1.6720  & 1.7730 & 94.30\% \\
   \hline  
\bottomrule          
\end{tabular}
}
\vspace{-10pt}
\end{table}

We aim to evaluate the quality of AI-generated items, comparing to real human-designed and photographed images.
Our human evaluation is conducted by a core panel of 8 experts, comprised of 6 professional fashion designers and 2 retail specialists, each with over 10 years of experience in the industry. 
To ensure quality and scale, annotations from a larger team of outsourced labelers are all reviewed and validated by these experts, guaranteeing a high standard of accuracy and professionalism across the entire evaluation dataset.
We assess the images from six aspects, including \textit{material quality}, \textit{model realism}, \textit{craft design}, \textit{pattern appropriateness}, \textit{design aesthetics}, and \textit{color coordination}. 
For each aspect, experts rate the images using a three-tier scoring system: 0 for poor, 1 for average, and 2 for good. 

The evaluation results are shown in Table~\ref{tab:human expert}. 
We can observe that AI-generated item images are generally close to human-designed real ones, especially in design aspects such as material quality, pattern appropriateness, and color coordination. 
While there is still room for improvement in craft design, our practical validation demonstrates that current AI-generated items (AIGI) can already support online e-commerce retail without the need for real-world manufacturing or physical prototypes beforehand.

\subsection{Product Lifecycle \& Manufacture Feasibility}

A primary concern for our ``sell it before you make it'' model is its real-world viability, specifically regarding the production lifecycle, manufacturing feasibility, and customer satisfaction. 
Our deployment at Alibaba has validated this innovative business mode by establishing a seamless and efficient workflow from digital design to physical product. The lifecycle and feasibility are detailed below.

\subsubsection{Product Lifecycle}
Our AIGI service is a merchant-facing tool, where end customers are generally \textbf{unaware} that a product originated from an AI design. This ensures a natural and uninterrupted shopping experience. The process unfolds in three key stages:
\begin{itemize}[leftmargin=10pt]
    \item \textit{AI-Assisted Design}: Merchants use our service to generate and select photorealistic designs to be listed in online stores.
    \item \textit{Demand Aggregation}: The AI-generated products appear alongside human-designed items on the platform. Customers browse and place orders as they normally would.
    \item \textit{Manufacturing}: Customer orders are aggregated. Once demand reaches a set threshold, a bulk production order is triggered. The manufactured products are then delivered to end customers.
\end{itemize}

\subsubsection{Manufacturing Feasibility \& Customer Satisfaction}
A core challenge for an AI-driven, on-demand fashion model is the feasibility of rapid, high-fidelity manufacturing. We validated our system through a large-scale industrial deployment, demonstrating that \textit{modern supply chains can support this approach at scale}.

Our primary validation comes from a \textbf{6-month} online A/B test across \textbf{17 product categories}. We compared the customer return rates of our AI-generated items against human-designed items within the same stores. The AI-generated products exhibited a \textbf{7.9\% lower return rate}, providing strong quantitative evidence of high customer satisfaction. This result indicates that the physical products successfully meet or exceed the expectations set by the AI-generated images.
The operational success of this model is underpinned by two key factors of manufacturing facilities:
\begin{itemize}[leftmargin=10pt]
    \item \textbf{Production Fidelity.} The AI-generated images serve as sufficient and direct blueprints for manufacturing, enabling accurate production without physical prototypes.
    \item \textbf{Rapid Turnaround.} Production cycles range from 3 to 20 days, with a mean of 7 days, making the on-demand customer experience competitive with traditional in-stock inventory models.
\end{itemize}
This confirms that our integration of generative AI and responsive manufacturing constitutes a viable and scalable paradigm.

\subsection{Post-Launch Performance}

\begin{table}
\vspace{-5pt}
\caption{The relative improvement of click-through rates (CTR), conversion rate and return rate of AI-generated items compared with their non-AI counterparts in the same store.}
\vspace{-10pt}
\label{tab:Rel.Imrpv}
\resizebox{0.48\textwidth}{!}{
\renewcommand\arraystretch{1.5}
\begin{tabular}{c|c|c|c|c}
\toprule
   \hline  
\multicolumn{1}{c|}{} & CTR for Search$\uparrow$ & CTR for recommendation$\uparrow$ & Conversion Rate$\uparrow$ & Return Rate$\downarrow$ \\ 
\hline 
Rel.Imrpv. & +17.81\% & +13.35\% & +17.27\% & -7.9\% \\
   \hline  
\bottomrule          
\end{tabular}
}
\vspace{-5mm}
\end{table}

\begin{figure*}
    \vspace{-5pt}
    \includegraphics[width=0.88\textwidth]{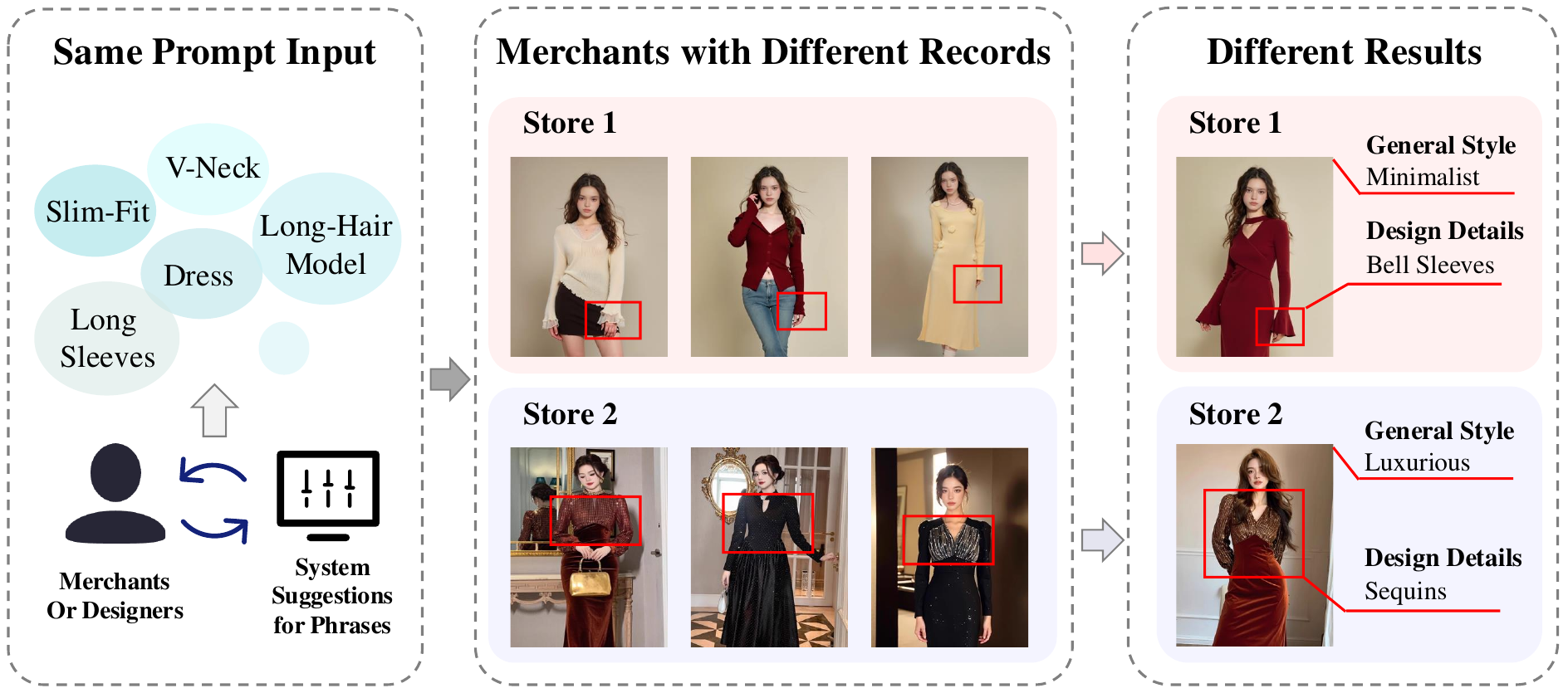}
  \vspace{-10pt}
  \caption{The case study of our deployed personalized text-to-image generation for AI-generated items (AIGI). 
  }
  \vspace{-10pt}
  \label{fig:case study}
\end{figure*}

We report relative improvements of click-through rate (CTR) for both search and recommendation scenarios, as well as conversion rate and return rate, for the AI-generated items compared with human-designed ones. 
The relative improvement is derived by comparing AI and non-AI items within the same e-commerce store and averaging the results at the store level. 
For fair comparison, AI-generated items and human-designed items are treated equally by the platform in terms of search and recommendation traffic. There is \textbf{no preferential treatment} towards either product type. 

The results are reported in Table~\ref{tab:Rel.Imrpv}. 
We observe that the click-through rate (CTR) for AI-generated items has increased by an average of \textbf{15.58\%} in both the search and recommendation scenarios, while the conversion rate has improved by \textbf{17.27\%}. 
This highlights the advantages of AI-generated items (AIGI) in attracting user clicks and driving purchase conversions.
The return rate for AI-generated items is also \textbf{7.9\% lower} than their human-designed counterparts in the same stores, indicating strong customer satisfaction with both product quality and delivery times. 
This success has fueled significant adoption. 
Over \textbf{6 months}, the number of stores using AIGI services has increased by \textbf{2.5$\times$}, with monthly GMV for AI products per store rising by \textbf{2.81$\times$}. The number of new AI products launched per month has grown by \textbf{2.71$\times$}.

We attribute these improvements to two potential factors. 
\textit{First}, the AI-generated digital models enhance the aesthetic appeal of the product images, which can be a challenge for smaller fashion shops. 
\textit{Second}, the rapid iteration capabilities of our AI-driven design process significantly reduce the time needed to launch new products from \textit{month-level} cycles to \textit{day-level} cycles. 
This acceleration eliminates the need for human design, prototype production, and model photography, allowing items to remain more closely aligned with fast-changing fashion trends and user preferences.

\subsection{Case Study}

This case study demonstrates how PerFusion captures personalized preferences across different merchants.
The construction of the text prompt in our application is simply concatenating different descriptive phrases with commas. 
As shown in Figure~\ref{fig:case study}, the designers from two different e-commerce stores submit the same text prompt for fashion product generation, \ie, \textit{``dress,v-neck,slim-fit,long sleeves,long-hair model''}. 
Note that during the training of our model, the descriptive phrases can be shuffled for data augmentation, and the prompt is thereby order-insensitive to phrases. 
From Figure~\ref{fig:case study}, we draw the following observations:
\begin{itemize}[leftmargin=10pt]
    \item The resulting images are both highly consistent with the input prompt, \ie, a long-hair model with a v-neck, slim-fit, long-sleeve dress. 
    This means that our model can satisfy the general user preference with high-quality fashion product photographs. 
    \item More importantly, even when the text prompt is the same for designers from two different e-commerce stores, the resulting images are tailored to reflect the distinct preferences and records of individual users, with a significant difference in general style and design details. 
    We give our analysis as follows:
    \begin{itemize}[leftmargin=10pt]
        \item For store 1, the design highlights a minimalist style, which is reflected in the clean lines and simplicity of the overall designs. 
        Moreover, as indicated by the red boxes, we can observe that the product is designed with the bell sleeve feature as a distinct design element, which is highly aligned with the store's previous records.
        \item For store 2, the general style of the resulting image focuses on a luxurious aesthetic, which is consistent with the previous interaction records. 
        Also, as denoted by the red boxes, sequins are added as a special design detail, which is one of the designer's classical choices for fashion design.
    \end{itemize}
\end{itemize}
This demonstrates that our model can not only satisfy the general aesthetic preference and text-image consistency, but also capture the nuanced personalized preferences across different individuals. 

\section{Conclusion}

This paper presents an innovative system deployed at Alibaba that leverages personalized text-to-image generation for AI-generated items (AIGI), enabling the ``sell it before you make it'' business mode. 
We propose a novel PerFusion framework to model the users' group-level personalized preferences under such scenarios. 
We conduct both offline and online experiments to validate the effectiveness of our algorithm. 
The AI-generated items have achieved over 13\% improvement in both click-through rate and conversion rate, as well as 7.9\% decrease in return rate, over human-designed ones, highlighting the revolutionary potential of our developed AIGI for e-commerce platforms.
Future work could explore more comprehensive personalization from merchants to end customers.


\begin{acks}
The Shanghai Jiao Tong University team is supported by National Natural Science Foundation of China (624B2096, 62322603).
This work was supported by Alibaba Group through Alibaba Research Intern Program.
\end{acks}

\bibliographystyle{ACM-Reference-Format}
\bibliography{sample-base}

\appendix

\section{Objective Derivation}
\label{app:loss derive}

\subsection{Ideal Preference Distribution}

The classical pairwise preference optimization algorithms such as DPO~\cite{rafailov2024direct,wallace2024diffusion} are derived based on the Bradley-Terry model.
In order to consider the preference optmization towards multiple candidates, we have to replace the Bradley-Terry model~\cite{bradley1952rank} with the Plackett-Luce model~\cite{luce1959individual,plackett1975analysis}. 
Specifically, the Plackett-Luce model probability for a ranking is given by:
\[
P(\sigma) = \prod_{i=1}^{n} \frac{\theta_{\sigma(i)}}{\sum_{j=i}^{n} \theta_{\sigma(j)}},
\]
where \( \sigma \) is a permutation representing the ranking of items, \( \theta_{\sigma(i)} \) is the score of the item in position \( i \), and \( \sum_{j=i}^{n} \theta_{\sigma(j)} \) is the sum of the scores of the remaining items. 

In our application, each item is the generated image $x$ for the text condition $c$. 
The score for each item can be expressed by the exponential over the reward function, \ie, $\exp(r(x,c))$. 
Given a candidate image set $\mathcal{X}$ divided into the positive set $\mathcal{P}$ and the negative set $\mathcal{N}$, the ideal ranking of these candidates is to rank every positive image before the negative ones. 
Hence, we make an independent assumption among positive samples, and derive the probability of ideal preference distribution as follows:
\begin{equation}
\begin{aligned}
    &p^{*}(x_p  \succ x_n,\forall x_p\in\mathcal{P},\forall x_n\in\mathcal{N}) \\
    =&\sum_{x_p\in\mathcal{P}}p(x_p  \succ x_n,\forall x_n\in\mathcal{N}) \;\;\;\;\;\;\;\; \text{(by independent assumption)} \\
    =& \sum_{x_p\in\mathcal{P}} \sum_{\sigma(1)=x_p} \prod_{i=1}^{|\mathcal{N}|+1}\frac{\exp\left(r(\sigma(i),c)\right)}{\sum
    _{j=i}^{|\mathcal{N}|+1} \exp\left(r(\sigma(j),c)\right)} \\
    =&\sum_{x_p\in\mathcal{P}}  \frac{\exp(r(x_p,c))}{\sum_{x^{\prime}\in\mathcal{N}\cup\{x_p\}}\exp(r(x^{\prime},c))} \sum_{\sigma \in \text{Perm}(\mathcal{N})}\prod_{i=1}^{|\mathcal{N}|}\frac{\exp\left(r(\sigma(i),c)\right)}{\sum_{j=i}^{|\mathcal{N}|}\exp\left(r(\sigma(j),c)\right)} \\
    =&\sum_{x_p\in\mathcal{P}}  \frac{\exp(r(x_p,c))}{\sum_{x^{\prime}\in\mathcal{N}\cup\{x_p\}}\exp(r(x^{\prime},c))} \\
    =&\sum_{x_p\in\mathcal{P}}\frac{1}{1+\sum_{x_n\in\mathcal{N}}\exp(r(x_n,c)-r(x_p,c))},
\end{aligned}
\label{app eq:preference distribution}
\end{equation}
where $\text{Perm}(\cdot)$ is full permutations of a given set, and, according to previous works~\cite{wallace2024diffusion,rafailov2024direct}, the reward $r(x,c)$ can be derived as:
\begin{equation}
    r(x,c)=\beta \log\frac{p_{\theta}(x|c)}{p_{ref}(x|c)}+\beta\log Z(c),
\end{equation}
where $p_{\theta}(\cdot)$ is the target model, $p_{ref}(\cdot)$ is the corresponding reference model, and $Z(c)$ is the partition function.

\subsection{Loss Function}

According to previous works on preference optimization~\cite{wallace2024diffusion,rafailov2024direct,bai2022training}, the training of the target model $p_{\theta}(\cdot)$ is to maximize the log likelihood of the ideal ranking for a given set of candidates. 
We can first derive the log likelihood of the ideal ranking based on the Plackett-Luce model:
\begin{equation}
\begin{aligned}
    &\log p^{*}(x_p  \succ x_n,\forall x_p\in\mathcal{P},\forall x_n\in\mathcal{N})\\
    =&\sum_{x_p\in\mathcal{P}}\frac{1}{1+\sum_{x_n\in\mathcal{N}}\exp(r(x_n,c)-r(x_p,c))} \\
    =&\sum_{x_p\in\mathcal{P}}\frac{1}{1+\sum_{x_n\in\mathcal{N}}\exp\left(\beta \log\frac{p_{\theta}(x|c)}{p_{ref}(x_n|c)}-\beta \log\frac{p_{\theta}(x|c)}{p_{ref}(x_p|c)}\right)} \\
    = &\sum_{x_p\in\mathcal{P}} \log\sigma\bigg(-\log\sum_{x_n\in\mathcal{N}}\exp\Big(\beta \log\frac{p_{\theta}(x_n|c)}{p_{ref}(x_n|c)}-\beta \log\frac{p_{\theta}(x_p|c)}{p_{ref}(x_p|c)}\Big) \bigg),
\end{aligned}
\end{equation}
where $\sigma(\cdot)$ is the sigmoid function. 
Therefore, the group-level loss function is written as:
\begin{equation}
\begin{aligned}
    \mathcal{L}=&-\mathbb{E}_{(u,c,\mathcal{P},\mathcal{N})}\left[\log p^{*}(x_p  \succ x_n,\forall x_p\in\mathcal{P},\forall x_n\in\mathcal{N})\right]\\
    = &-\mathbb{E}_{(u,c,\mathcal{P},\mathcal{N})}\Bigg[\sum_{x_p\in\mathcal{P}} \log\sigma\bigg(-\log\sum_{x_n\in\mathcal{N}}\exp\Big(\\
    &\;\;\;\;\;\;\;\;\;\beta \log\frac{p_{\theta}(x_n|c)}{p_{ref}(x_n|c)}-\beta \log\frac{p_{\theta}(x_p|c)}{p_{ref}(x_p|c)}\Big) \bigg)
    \Bigg].
\end{aligned}
\end{equation}
Note that it is intractable to calculate the likelihood of an image, \ie, $p_{\theta}(x|c)$. Therefore, following previous work~\cite{wallace2024diffusion}, we approximate the denoising trajectory $p_{\theta}(x|c)$ with the diffusion process $q(x|c)$ and derive the final objective:
\begin{equation}
\small
    \mathcal{L}= -\mathbb{E}_{(u,t,c,\mathcal{P},\mathcal{N})} \sum_{x_p\in\mathcal{P}} \log\sigma\bigg(-\log\sum_{x_n\in\mathcal{N}}\exp\Big(\beta s(x_p,c,t,u)-\beta s(x_n,c,t,u)
    \Big) \bigg),
\label{app eq:group loss}
\end{equation}
where $s(x,c,t,u)=\|\epsilon^* - \epsilon_\theta(z_t,t,c,u)\|^2_2 - \|\epsilon^* - \epsilon_\text{ref}(z_t,t,c,u)\|^2_2$, $u$ is the user preference representation, $z_t$ is the latent diffused representation of image $x$, and $\beta$ is a hyper-parameter for regularization. 

\subsection{Relation With DPO}

When we only have one positive image and one negative image, \ie, $|\mathcal{P}|=|\mathcal{N}|=1$, the training objective in Eq.~\ref{app eq:group loss} can be reduced to the classical direct preference optimization (DPO) objective for diffusion models proposed by \citet{wallace2024diffusion}. 
The derivation is as follows:
\begin{equation}
\small
\begin{aligned}
    \mathcal{L}= &-\mathbb{E}_{(u,t,c,\mathcal{P},\mathcal{N})} \sum_{x_p\in\mathcal{P}} \log\sigma\bigg(-\log\sum_{x_n\in\mathcal{N}}\exp\Big(\beta s(x_p,c,t,u)-\beta s(x_n,c,t,u)
    \Big) \bigg) \\
    =& -\mathbb{E}_{(u,t,c,\mathcal{P},\mathcal{N})} \log\sigma\left(-\log\ \exp\left(\beta s(x_p,c,t,u)-\beta s(x_n,c,t,u)
    \right) \right) \\ 
    =& -\mathbb{E}_{(u,t,c,\mathcal{P},\mathcal{N})} \log\sigma\left(\beta s(x_n,c,t,u)-\beta s(x_p,c,t,u)
    \right).
\end{aligned}
\end{equation}
Therefore, we can observe that the classical Diffusion DPO loss is a special case of our proposed group-level preference optimization objective. 
Our derived objective is more generalized for scenarios where we have to align preferences among multiple candidates.

\section{Experiment Setups}
\subsection{Baseline Models}
\label{app:baseline models}

For \textit{user preference estimation} task, we adopt Aesthetics~\cite{aesthetic_classifier}, CLIP-H~\cite{radford2021learning}, ImageReward~\cite{xu2024imagereward}, HPSv1~\cite{wu2023human}, HPSv2~\cite{wu2023humanv2} and PickScore~\cite{kirstain2023pick} as the baseline models:
\begin{itemize}[leftmargin=10pt]
    \item \textbf{Aesthetic} evaluates the visual appeal of generated images. The goal is to predict how aesthetically pleasing an image is, which often involves factors like color harmony, composition, sharpness, and overall visual quality. This type of model typically does not directly consider user preferences or image-text alignment but focuses on the intrinsic beauty of the image itself.
    \item \textbf{CLIP-H}
     evaluates the alignment between images and texts using the CLIP model. CLIP-H is trained to connect images and their textual descriptions by mapping both to a shared embedding space. CLIP-H specifically measures how well the content of an image corresponds to its textual description, \ie, the relevance and consistency of the text-image pair.
    \item \textbf{ImageReward}
    combines the factors of both aesthetics and human preference. It not only evaluates how visually appealing an image is (like the Aesthetic model) but also considers how well the image corresponds to human preferences or expectations.
    \item \textbf{HPSv1}
    works as a human preference classifier for images based on visual features such as texture, color, and other perceptual aspects. This model assesses how visually compelling or suitable an image is in relation to general human preference patterns.
    \item \textbf{HPSv2}
    is an improved version of HPSv1 and is designed to make more accurate human preference predictions. 
    \item \textbf{PickScore} models the general human preference by conducting pairwise training between a pair of positive and negative images for a given text prompt.
\end{itemize}
For \textit{personalized image generation} task, we adopt Stable Diffusion v1.5~\cite{rombach2022high}, SFT~\cite{ho2020denoising}, DPO~\cite{wallace2024diffusion}, SPO~\cite{liang2024step} and PMG~\cite{shen2024pmg} as baselines:
\begin{itemize}[leftmargin=10pt]
    \item \textbf{Stable Diffusion v1.5}
    is a pretrained generative model for creating high-quality images. 
    This model serves as not only a high-quality baseline for our personalized image generation task, but also the backbone for the other preference alignment algorithms.
    \item \textbf{SFT}
    stands for supervised finetuning with the pointwise diffusion model training objective as introduced in Eq.~\ref{eq:diffusion loss}.
    \item \textbf{SPO}
    (Stepwise Preference Optimization) improves image aesthetics by focusing on fine-grained visual differences at each denoising step of the generative process. 
    \item \textbf{DPO}
    (Direct Preference Optimization) conducts pairwise preference learning from human's binary feedback over a pair of images towards the given text prompt.
    \item \textbf{PMG} is a personalized multi-modal content generation framework by transforming user interaction records into textual descriptions and key phrases.
    
\end{itemize}

\subsection{Evaluation Metrics}
\label{app:evaluation metrics}

For the \textit{user preference estimation} task, which is formulated as a ranking task, we adopt the mean average precision (MAP) and group-wise area under the curve (GAUC) as our metrics:
\begin{itemize}[leftmargin=10pt]
    \item \textbf{MAP}
    measures the precision at different levels of the ranked list, averaged across all user interaction records. 
    A higher MAP score means that the model is better at prioritizing images that align with user preferences. 
    \item \textbf{GAUC} first calculates the AUC metric for each group (\ie, one user interaction record) and then produces the GAUC by averaging on the user interactions.
    A higher GAUC score implies better ranking quality.
\end{itemize}
For the \textit{personalized image generation} task, we adopt Aesthetic~\cite{aesthetic_classifier}, CLIP Score~\cite{radford2021learning}, PickScore~\cite{kirstain2023pick}, and PerFusionRM as our metrics.
\begin{itemize}[leftmargin=10pt]
    \item \textbf{Aesthetic}
    evaluates the visual appeal of the generated images, with higher values indicating better aesthetic quality.  
    \item \textbf{CLIP Score}
    measures consistency for a text-image pair. 
    A higher CLIP score implies that the generated images better match the textual inputs. 
    \item \textbf{PickScore}
    evaluates the relevance of generated images to general human preferences based on the text prompts. 
    Higher scores indicate that the generated images are more likely to be preferred by humans. 
    \item \textbf{PerFusionRM} is proposed in this paper to estimate the personalized user preference for a given text-image pair. 
    Hence, we adopt it to evaluate how well the generated image aligns with a specific user's preference. 
    A higher score indicates better personalized alignment for different users. 
\end{itemize}

\subsection{Implementation Details}
\label{app:implementation}

For PerFusionRM on the user preference estimation task, we use PickScore~\cite{kirstain2023pick} as the backbone model, which is an improved version of CLIP to better capture the general human preference. 
We adopt AdamW~\cite{loshchilov2017decoupled} as the optimizer. 
The batch size is set to 128. 
The learning rate is set to 1e-3 with a 500-step warmup. 
The weight decay is set to 1e-2.
PickScore and our PerfusionRM are tuned on both datasets, while other baselines stay untuned to capture different aspects of user vision preferences, e.g., aesthetics and text-image alignment.

For the personalized image generation task, we first conduct direct preference optimization (DPO)~\cite{wallace2024diffusion} on Stable Diffusion v1.5 to capture the general user preference, which is then adopted as the backbone model for our PerFusion. 
We train the model at fixed square
resolutions. 
We adopt AdamW~\cite{loshchilov2017decoupled} as the optimizer. 
We set the learning rate to 1e-8 with a 500-step warmup. 
The weight decay is set to 1e-2. 
The batch size is selected from $\{256, 512, 1024\}$. 
The hyperparameter $\beta$ is chosen from $\{1500, 2000, 2500, 3000\}$.
All baseline models use the same input prompts for fair comparison, except PMG, which is a method to improve input prompts based on user profiles.

\end{document}